# Deep Learning-enabled Detection and Classification of Bacterial Colonies using a Thin Film Transistor (TFT) Image Sensor

Yuzhu Li[†,1,2,3], Tairan Liu[†,1,2,3], Hatice Ceylan Koydemir[4,5], Hongda Wang[1,2,3], Keelan O'Riordan[6], Bijie Bai[1,2,3], Yuta Haga[7], Junji Kobashi[7], Hitoshi Tanaka[7], Takaya Tamaru[7], Kazunori Yamaguchi[7] and Aydogan Ozcan[*,1,2,3,8]

[1]Electrical and Computer Engineering Department, University of California, Los Angeles, CA, 90095, USA.
[2]Bioengineering Department, University of California, Los Angeles, 90095, USA.
[3]California NanoSystems Institute (CNSI), University of California, Los Angeles, CA, 90095, USA.
[4]Department of Biomedical Engineering, Texas A&M University, College Station, TX, 77843, USA.
[5]Center for Remote Health Technologies and Systems, Texas A&M University, College Station, TX, 77843, USA.
[6]Department of Physics and Astronomy, University of California, Los Angeles, CA, 90095, USA.
[7]Device Development Department, Research & Development Division, Japan Display Inc., Japan.
[8]Department of Surgery, University of California, Los Angeles, CA, 90095, USA.

*Correspondence: Aydogan Ozcan. Email: ozcan@ucla.edu

[†] Equal contributing authors

## Abstract

Early detection and identification of pathogenic bacteria such as *Escherichia coli* (*E. coli*) is an essential task for public health. The conventional culture-based methods for bacterial colony detection usually take ≥24 hours to get the final read-out. Here, we demonstrate a bacterial colony-forming-unit (CFU) detection system exploiting a thin-film-transistor (TFT)-based image sensor array that saves ~12 hours compared to the Environmental Protection Agency (EPA)-approved methods. To demonstrate the efficacy of this CFU detection system, a lensfree imaging modality was built using the TFT image sensor with a sample field-of-view of ~10 cm$^2$. Time-lapse images of bacterial colonies cultured on chromogenic agar plates were automatically collected at 5-minute intervals. Two deep neural networks were used to detect and count the growing colonies and identify their species. When blindly tested with 265 colonies of *E. coli* and other coliform bacteria (i.e., *Citrobacter* and *Klebsiella pneumoniae*), our system reached an average CFU detection rate of 97.3% at 9 hours of incubation and an average recovery rate of 91.6% at ~12 hours. This TFT-based sensor can be applied to various microbiological detection methods. Due to the large scalability, ultra-large field-of-view, and low cost of the TFT-based image sensors, this platform can be integrated with each agar plate to be tested and disposed of after the automated CFU count. The imaging field-of-view of this platform can be cost-effectively increased to >100 cm$^2$ to provide a massive throughput for CFU detection using, e.g., roll-to-roll manufacturing of TFTs as used in the flexible display industry.

# Introduction

Bacterial infection has been a leading factor that causes millions of deaths each year in both developed and developing countries[1,2]. The associated expenses of treating bacterial infections cost more than 4 billion dollars annually in the United States (US) alone[3]. Therefore, the rapid and accurate detection of pathogenic bacteria is of great importance to human health in preventing such infectious diseases caused by e.g., contamination in food and drinking water. Among those pathogenic bacteria, *Escherichia coli* (*E. coli*) and other coliform bacteria are among the most common ones, and they indicate fecal contamination in food and water samples[1]. The most basic and frequently used method of detecting *E. coli* and total coliform bacteria involves culturing the sample on a solid agar plate or liquid medium following the US Environmental Protection Agency (EPA)-approved protocols (e.g., EPA 1103.1 and EPA 1604 methods)[4,5]. However, these traditional culture-based methods usually take ≥24 hours for the final read-out and need visual recognition and counting of colony-forming units (CFUs) by microbiology experts. Although various nucleic acid-based molecular detection approaches[6–9] have been developed for rapid bacteria detection with results ready in less than a few hours, they present lower sensitivity in general and have challenges to differentiate live and dead bacteria[10]; in fact, there is no EPA-approved nucleic acid-based coliform sensing method that can be used for screening water samples. Various other approaches have been developed to provide high sensitivity and specificity for the detection of bacteria based on different methods such as e.g., fluorimetry[11], solid-phase cytometry[12], fluorescence microscopy[13], Raman spectroscopy[14] and others[15–17]; however, these systems, in general, do not work with large sample volumes (e.g., >0.1 L). As another alternative, Wang *et al.* demonstrated a complementary metal-oxide-semiconductor (CMOS) image sensor-based time-lapse imaging platform to perform early detection and classification of coliform bacteria[10]. This method achieved more than 12 hours of detection time savings and provided species classification with >80% accuracy within 12-hours of incubation. The field-of-view (FOV) of the CMOS image sensor in this design was < 0.3 cm$^2$, and therefore the mechanical scanning of the Petri dish area was required to obtain an image of the whole FOV of the cultured sample. Not only that this is time-consuming and requires additional sample scanning hardware, but it also brings some extra digital processing burden for image registration and stitching.

Recently, with the fast development of thin-film-transistors (TFT), the TFT technology has been widely used in the field of flexible display industry[18], radio frequency identification tags[19], ultrathin electronics[20], and large-scale sensors[21,22] thanks to its high scalability, low-cost mass production (involving e.g., roll-to-roll manufacturing), low power consumption, and low heat generation properties. TFT technology has also been applied in the biosensing field to detect pathogens by transferring e.g., antibody-antigen binding, enzyme-substrate catalytic activity, or DNA hybridization into electrical signals[22,23]. For example, a low-cost TFT nanoribbon sensor was developed by Hu *et al.* to detect the gene copies of *E. coli* and *Klebsiella pneumoniae (K. pneumoniae)* in a few minutes by using PH change due to DNA amplification[24]. As another example, Salinas *et al.* implemented a ZnO TFT biosensor with recyclable plastic substrates for real-time *E. coli* detection[25,26]. However, these TFT-based biosensing methods could not differentiate between live and dead bacteria and did not provide quantification of the CFU concentration of the sample under test.

Here, we demonstrate the first use of a TFT-based image sensor to build a real-time CFU detection system to automatically count the bacterial colonies and rapidly identify their species using deep learning. Because of the large FOV of the TFT image sensor (~10 cm$^2$), there is no need for mechanical scanning of the agar plate, which enabled us to create a field-portable and cost-effective lensfree CFU detector as shown in Fig. 1. This compact system includes sequentially switched red, green, and blue light-emitting diodes (LEDs) that periodically illuminate the cultured samples (*E. coli*, *Citrobacter,* and *K. pneumoniae*) as shown in Fig. 1(c), and the spatio-temporal patterns of the samples are collected by the TFT image sensor, with an imaging period of 5 min. Two deep learning-based classifiers were trained to detect the bacterial colonies and then classify them into *E. coli* and total coliform bacteria. Blindly tested on a dataset populated with 265 colonies (85 *E. coli* CFU, 66 *Citrobacter* CFU, and 114 *K. pneumoniae* CFU), our TFT-based system was able to detect the presence of the colonies as early as ~6 hours during the incubation period and achieved an average CFU detection rate of 97.3% at 9 hours of incubation, saving more than 12 hours compared to the EPA-approved culture-based CFU detection methods. For the classification of the detected bacterial colonies, an average recovery rate of 91.6% was achieved at ~12 hours of incubation.

This TFT-based field-portable CFU detection system significantly benefits from the cost-effectiveness and ultra-large FOV of TFT image sensors, which can be further scaled up, achieving even lower costs with much larger FOVs based on e.g., roll-to-roll manufacturing methods commonly used in the flexible display industry. We believe that TFT image sensors can potentially be integrated with each agar plate to be tested, and can be disposed of after the determination of the CFU count, opening up various new opportunities for microbiology instrumentation in the laboratory and field settings.

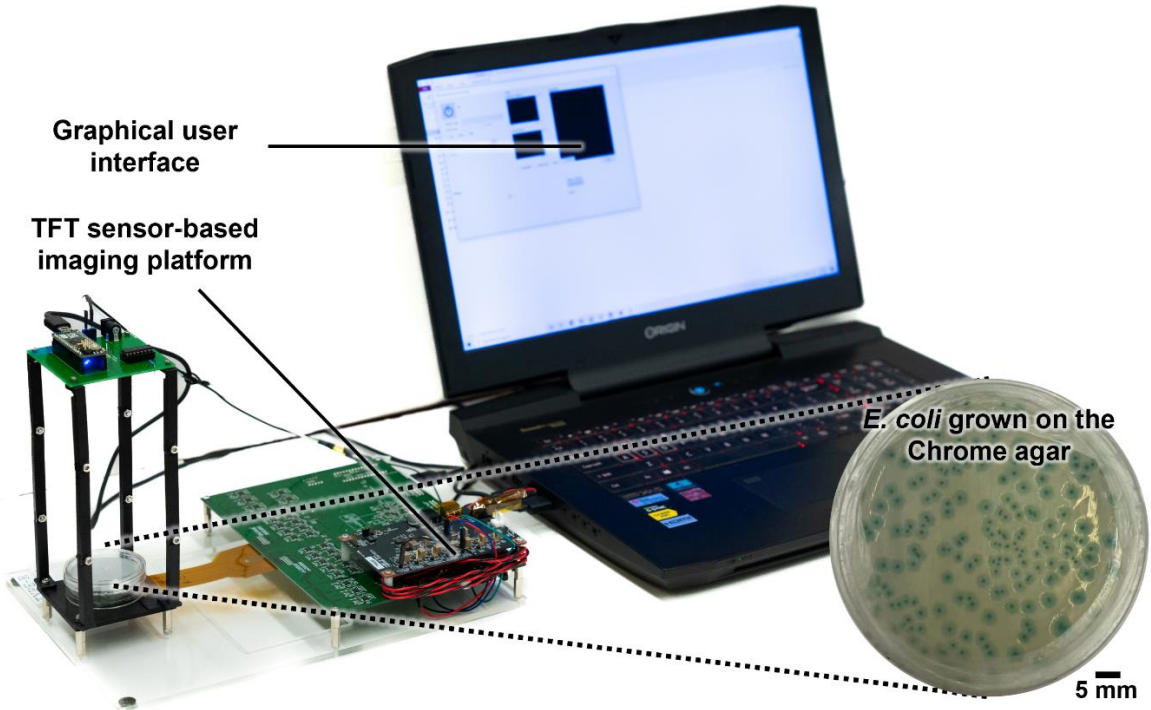

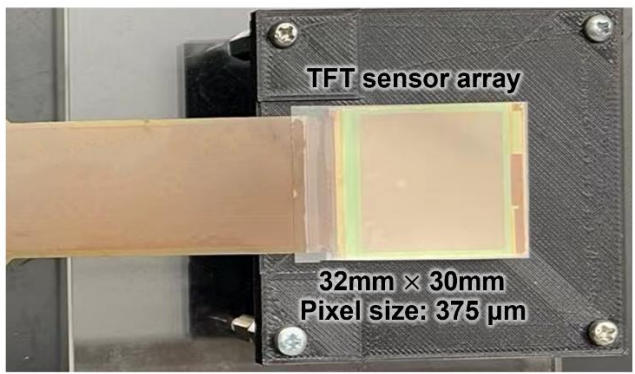

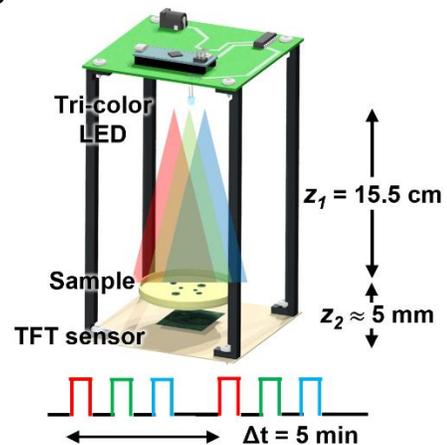

**Fig. 1: Real-time CFU detection and classification system using a TFT image sensor.** (a) A photo of the lensfree imaging system, samples to be tested, and the laptop computer used for controlling the hardware. The chromogenic agar medium results in a gray-green color for *E. coli* colonies and a pinkish color for other coliform bacteria; furthermore, it inhibits the growth of different bacterial colonies or exhibits colorless colonies when other types of bacteria are present in the sample. (b) A zoomed-in photo of the TFT image sensor with a FOV of 32 mm × 30 mm. (c) Detailed illustration of the lensfree imaging modality. The red (620 nm), green (520 nm), and blue (460 nm) LEDs were switched on sequentially at 5-minute intervals to directly illuminate the cultured

samples, which were imaged by the TFT image sensor in a single shot. The distance between the tri-color LED and the agar plate sample ($z_1$) is 15.5 cm, while the sample to sensor distance ($z_2$) distance is ~5 mm.

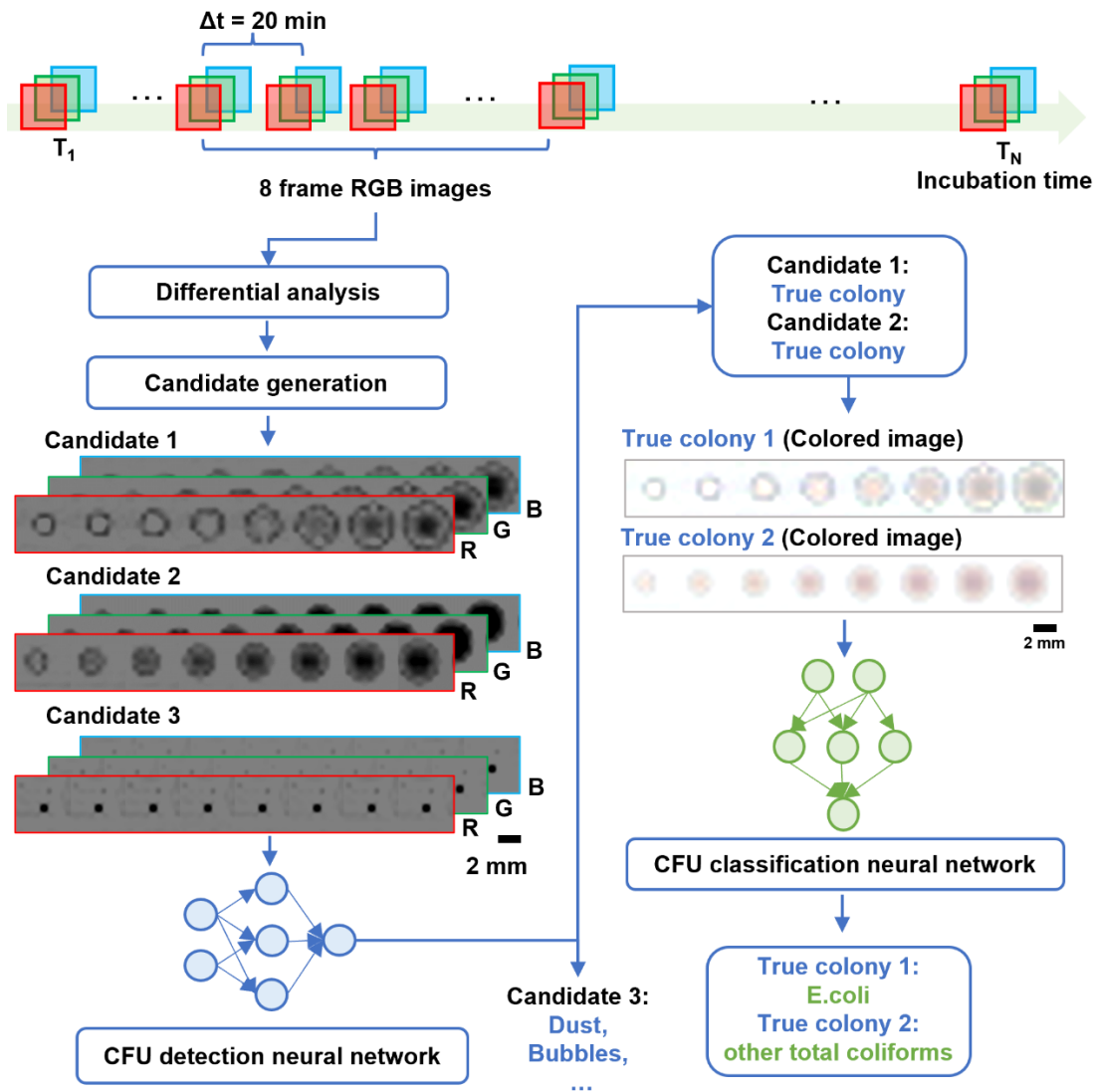

**Fig. 2: Schematics of the workflow of our deep learning-based CFU detection and classification system.** We process 8 whole FOV RGB images with 20-minute time intervals for the differential analysis to select the initial colony "candidates". The digitally-cropped 8-frame RGB image sequence for each individual colony candidate is fed into the CFU detection neural network first. This neural network rejects various non-colony objects (among the initial colony candidates) such as dust and bubbles, achieving true colony detection. Next, the detected colonies are passed through the CFU classification neural network to identify their species (*E. coli* or other total coliforms, i.e., binary classification).

## Results

We experimentally demonstrated the success of our framework by detecting and classifying the colonies *E. coli* and two other types of total coliform bacteria, i.e., *Citrobacter* and *K. pneumoniae,* on chromogenic agar plates, which result in a gray-green color for *E. coli* colonies and a pinkish color for other coliform bacteria, also inhibiting the growth of different bacterial colonies when other types of bacteria exist in the sample. Each sample was prepared following the EPA-1103.1 method[5] (see the Methods section) using a Petri dish. After the sample was prepared, it was directly placed on top of the TFT image sensor as part of the lensfree imaging system, and the entire imaging modality (except the laptop in Fig. 1(a)) was placed inside an incubator to record the growth of the colonies with 5-minute imaging intervals. For each time interval, three images were collected sequentially using the TFT image sensor under red (620 nm), green (520 nm), and blue (460 nm) illumination light. This multi-wavelength design allowed the monochromatic TFT image sensor to reconstruct color images of the bacterial colonies and was mainly used to identify their species by exploiting the color information provided by the selective chromogenic agar medium. The recorded time-lapse images were processed using the workflow shown in Fig. 2, where a differential analysis was used to select the initial colony candidates, and two deep neural networks (DNNs) were trained to further screen the colony candidates to specifically detect the true colonies and infer their species (see the Methods section for details). All these image processing steps take <25 sec using an Intel Core i7-7700 CPU-powered computer, consuming <1 GB of memory (without the need for GPUs).

The presented TFT imaging system periodically captures the images of the agar plate under test based on lensfree in-line holography; however, due to its large pixel size (375 μm) and relatively small sample to sensor distance (~5 mm, which is equal to the thickness of the agar), a free space backpropagation[27–31] step is not needed. By directly using the raw intensity images as part of the RGB color channels and calibrating the background, the color images of the agar plate can be generated in <0.25 sec after the TFT images are recorded. Figure 3 shows examples of color images of *E. coli*, *Citrobacter,* and *K. pneumoniae* colonies at different stages of their growth, captured by our system (also see Videos S1-S3). Consistent with the EPA-approved method (EPA-1103.1[5]), *E. coli* colonies exhibit gray-green colors, while *Citrobacter* and *K. pneumoniae* colonies exhibit pinkish color using the chromogenic agar.

Based on the imaging performance of our TFT-based CFU detection system summarized in Fig.3, we quantified its early detection and classification performance as shown in Fig.4. For this, we trained the detection and the classification neural network models (see the Methods section for training details) on a dataset of 442 colonies (128 *E. coli* colonies, 126 *Citrobacter,* and 188 *K. pneumoniae* colonies) captured from 17 independent experiments. The testing dataset was populated using 265 colonies from 13 independent experiments, which had a total of 85 *E. coli* colonies, 66 *Citrobacter* colonies, and 114 *K. pneumoniae* colonies. The detection rate was defined as the ratio of the number of true colonies confirmed by the CFU detection neural network out of the total colony number counted by an expert after 24-hour incubation. Figure 4(a, c, e) shows the detection rate we achieved in the blind testing phase as a function of the incubation time. As shown in Fig. 4(a, c, e), > 90% detection rate was achieved at 8 hours of incubation for *E. coli*, 9 hours for *Citrobacter*, and 7 hours 40 minutes for *K. pneumoniae*. Furthermore, a 100% detection rate was obtained within 10 hours of incubation for *E. coli*, 11 hours for *Citrobacter*, and 9 hours 20 minutes for *K. pneumoniae*. Compared to the EPA-approved standard read-out time (24 hours), our TFT-based CFU detection system achieved > 12 hours of time-saving. Moreover, from the detection rate curves reported in Fig. 4, we can also qualitatively infer that the colony growth speed of *K. pneumoniae* is larger than *E. coli* which is larger than *Citrobacter* because the earliest detection times for *E. coli*, *Citrobacter,* and *K. pneumoniae* colonies were 6 hours, ~6.5 hours and ~5.5 hours of incubation, respectively.

To quantify the performance of our bacterial colony classification neural network, the recovery rate was defined as the ratio of the number of correctly classified colonies to the total number of colonies counted by an expert after 24-hour incubation. Figure 4(b, d, f) shows the recovery rate curves over all the blind testing experiments as a function of the incubation time. We can see that a recovery rate of > 85% was achieved at 11 hours 20 minutes for *E. coli*, at 13 hours for *Citrobacter*, and at 10 hours 20 minutes for *K. pneumoniae.* It is hard to achieve a 100% recovery rate for all the colonies since some of the late growing "wake-up" colonies could not grow to a sufficiently large size with the correct color information even after 24 hours of incubation. Fig. 4 also reveals that there exists approximately a 3-hour time delay between the colony detection time and species identification time; this time delay is expected since more time is needed for the detected colonies to grow larger and provide discernable color information for the correct classification of their species.

## Discussion

Note that our presented results in Fig. 4 represent a conservative performance of our TFT-based CFU detection method since the ground truth colony information was obtained after 24 hours of incubation. In the early stages of the incubation period, some bacterial colonies did not even exist physically. Therefore, if we used the existing colony numbers for each time point as the ground truth, we would have reported even higher detection and recovery rates in Fig. 4.

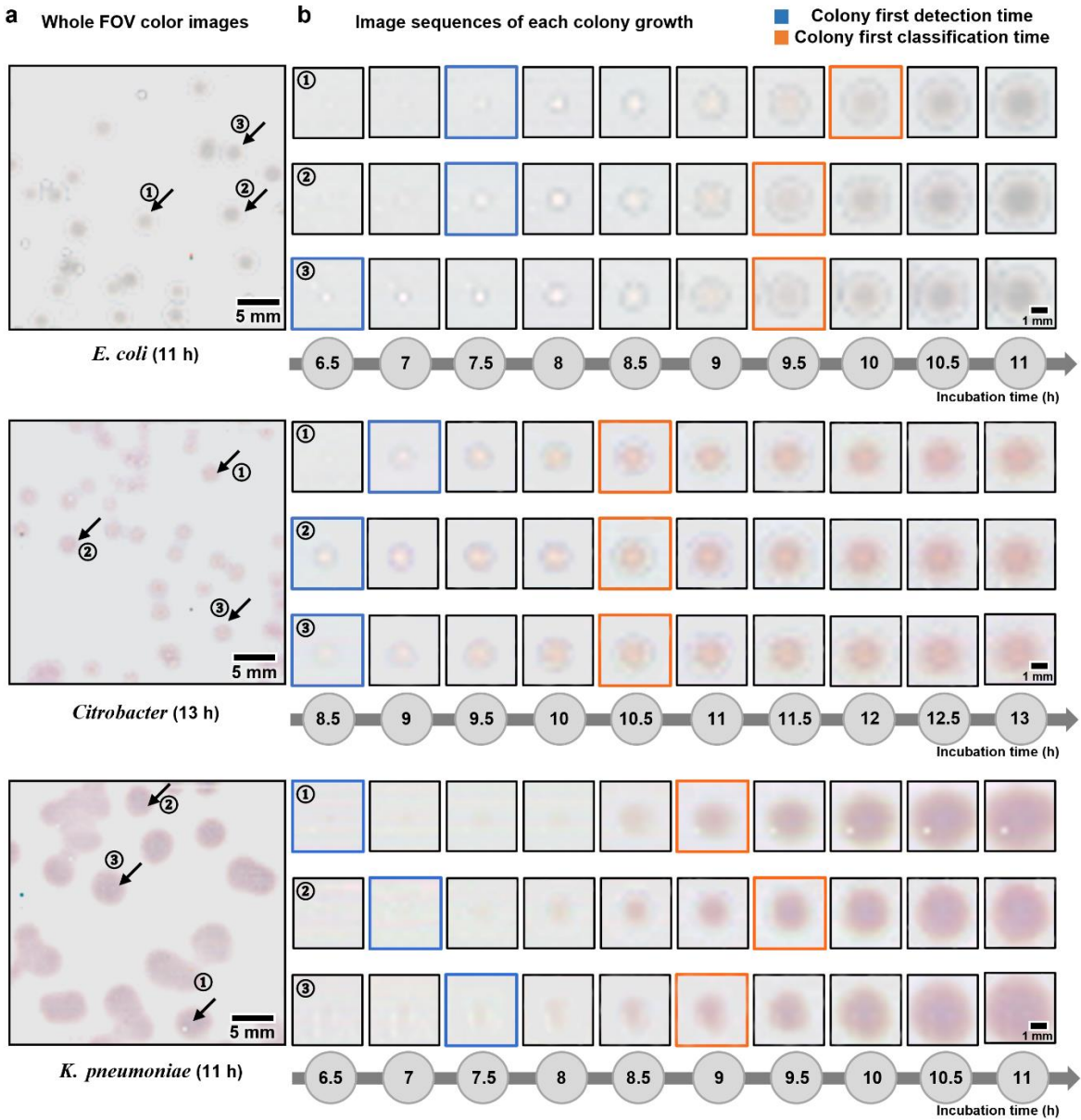

**Fig. 3: Visual evaluation of coliform bacterial colony early detection and classification using a TFT image sensor.** (a) Whole FOV color images of *E. coli* at 11-hour incubation, *Citrobacter* at 13-hour incubation, and *K. pneumoniae* at 11-hour incubation. (b) Examples of the image sequence of each isolated colony growth. Three independent colony growth sequences were selected for each one of the bacteria species. The blue box labels the first colony detection time confirmed by the CFU detection neural network, and the orange box corresponds to the first classification time correctly predicted by the CFU classification neural network.

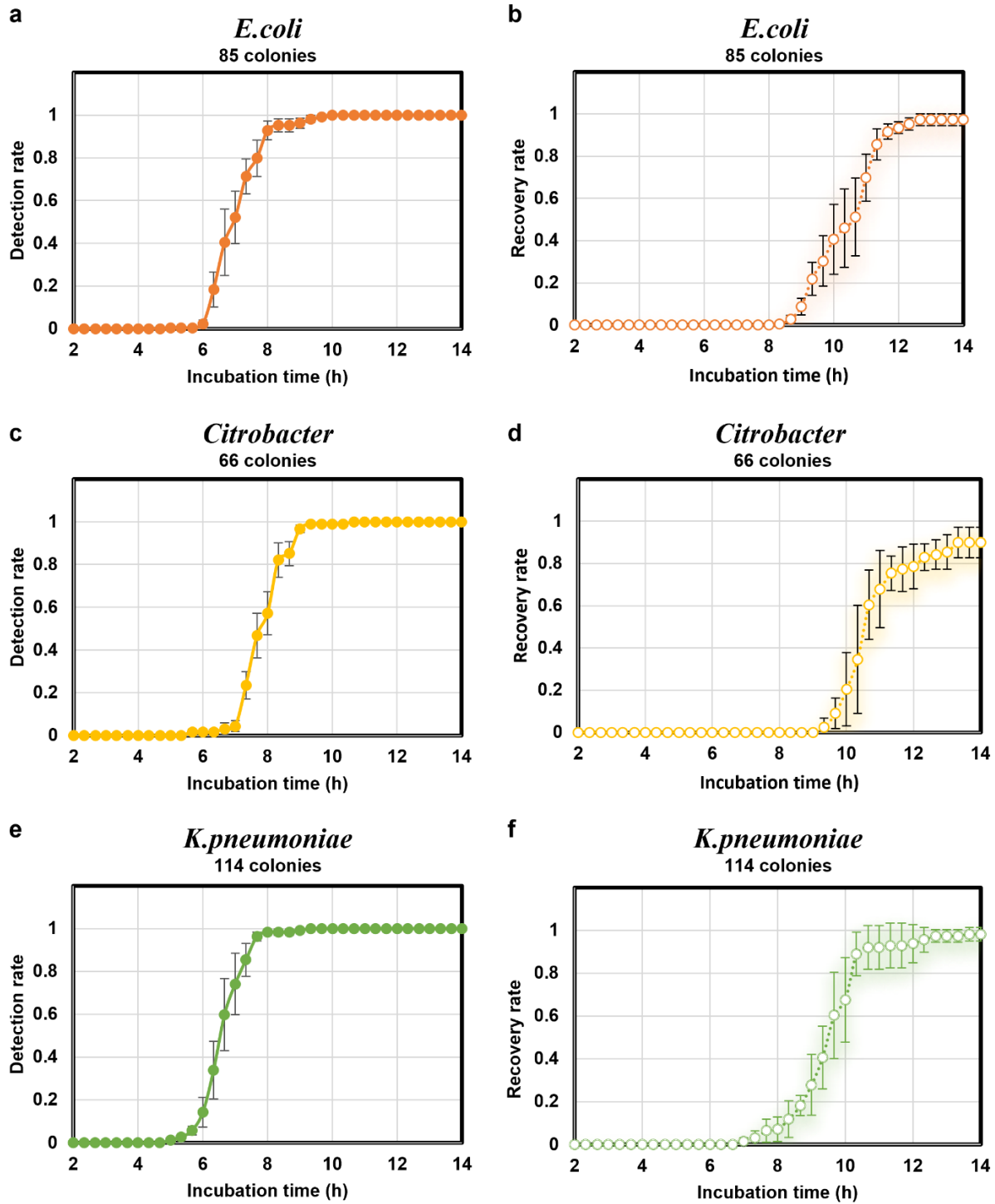

**Fig. 4: Quantitative performance evaluation of coliform colony early detection and classification using a TFT image sensor.** (a, c, e) The colony detection rate as a function of the incubation time for *E. coli*, *Citrobacter,* and *K. pneumoniae*. The mean and standard deviation of the detection rate were calculated on 85 *E. coli* colonies, 66 *Citrobacter* colonies, and 114 *K. pneumoniae* colonies for each time point. (b, d, f) The colony recovery rate as a function of the incubation time for *E. coli*, *Citrobacter,* and *K. pneumoniae*. The mean and standard deviation of the recovery rate were calculated on 85 *E. coli* colonies, 66 *Citrobacter* colonies, and 114 *K. pneumoniae* colonies for each time point.

Overall, the performance of our TFT-based CFU detection method is similar to the CMOS-based time-lapse imaging method[10] in terms of the colony detection speed. However, due to its large pixel size (375 μm) and limited spatial resolution, the TFT-based method has a slightly delayed colony classification time. With its ultra-large imaging FOV (~10 cm$^2$), the TFT-based CFU detection method eliminates (1) the time-consuming mechanical scanning of the Petri dish and the related optomechanical hardware, and (2) the image processing steps for image registration and stitching that would both be required due to the limited FOV of CMOS-based imagers. In addition to saving image processing time, this also helps the system to increase the CFU detection sensitivity as the system is free from any image registration and stitching artifacts and therefore, it can precisely capture minute spatio-temporal changes in the agar caused by bacterial colony growth at an early stage. Due to the massive scalability of the TFT arrays, the imaging FOV of our platform can be further increased to several tens to hundreds of cm$^2$ in a cost-effective manner, which could provide unprecedented levels of imaging throughput for automated CFU detection using e.g., roll-to-roll manufacturing of TFTs, as employed in the flexible display industry.

Another prominent advantage of the TFT-imager based detection system is that it can be adapted to image a wide range of biological samples using cost-effective and field-portable interfaces. Should the users have any contamination concerns, the TFT image sensor shown in Fig. 1 can be replaced and even used in a disposable manner (e.g., integrated as part of the Petri dish). Furthermore, the heat generated by the TFT image sensor during the data acquisition process is negligible, ensuring that the biological samples can grow at their desired temperature without being perturbed. Finally, our TFT-based CFU detection system is user-friendly and easy-to-use because there is no need for complex optical alignment, high precision mechanical scanning stages, or image registration/alignment steps.

In summary, we believe that the presented CFU detection system using TFT image sensor arrays provides a high-throughput, cost-effective, and easy-to-use solution to perform early detection and classification of bacterial colonies, opening up unique opportunities for microbiology instrumentation in the laboratory and field settings.

## Materials and Methods

### Sample preparation

All the bacterial sample preparations were performed at our Biosafety Level 2 laboratory in accordance with the environmental, health, and safety rules of the University of California, Los Angeles. We used *E. coli* (Migula) Castellani and Chalmers (ATCC® 25922™), *Citrobacter* (ATCC® 43864™), and *K. pneumoniae* subsp. *pneumoniae* (Schroeter) Trevisan (ATCC®13883™) as our culture microorganisms. CHROMagar™ ECC (product no. EF322, DRG International, Inc., Springfield, NJ, USA) chromogenic substrate mixture was used as the solid growth medium to detect *E. coli* and other total coliform colonies.

For each time-lapse imaging experiment, a bacterial suspension in a phosphate-buffered solution (PBS) (product no. 20-012-027, Fisher Scientific, Hampton, NH, USA) was prepared from a solid agar plate incubated for 24 hours. The concentration of the suspension was measured using a spectrophotometer (model no. ND-ONE-W, Thermo Fisher). Then, a serial dilution was performed in PBS to finally reach a concentration of ~10$^3$ CFUs / mL. Around 100 μL diluted suspension with ~100 CFUs was spread on a CHROMagar™ ECC plate using an L-shaped spreader (product no. 14-665-230, Fisher Scientific, Hampton, NH, USA). Next, the plate was covered with its lid, inverted, and placed on the TFT image sensor, which was placed with the whole imaging system into an incubator (product no. 151030513, ThermoFisher Scientific, Waltham, MA, USA) kept at 37 ± 0.2 °C.

Additionally, CHROMagar™ ECC plates were prepared ahead of time using the following method. CHROMagar™ ECC (6.56 g) was mixed with 200 mL of reagent grade water (product no. 23-249-581, Fisher Scientific, Hampton, NH, USA). The mixture was then heated to 100 °C on a hot plate while being stirred regularly using a magnetic stirrer bar. After cooling the mixture to ~50 °C, 10 mL of the mixture was dispensed into each Petri dish (60 mm × 15 mm) (product no. FB0875713A, Fisher Scientific, Hampton, NH, USA). When the agar plates solidified, they were sealed using parafilm (product no. 13-374-16, Fisher Scientific, Hampton, NH, USA), and covered with aluminum foil to keep them in the dark before use. These plates were stored at 4 °C and were used within two weeks after preparation.

**Imaging Set-up**

Our field-portable CFU imager comprises an illumination module and a TFT-based image sensor. The light from a tri-color LED directly illuminates the samples and forms in-line holograms on the TFT image sensor (JDI, Japan Display Inc., Japan). The TFT module includes a controlling printed circuit board (PCB) that provides the illumination and image capture control signal and an image sensor (with 80×84 pixels, pixel size = 375 μm). For the illumination module, a tri-color LED (EDGELEC) was controlled by a microcontroller (Arduino Micro, Arduino LLC) through a constant current LED driver (TLC5916, Texas Instrument, TX, USA) to sequentially provide the red (620 nm), green (520 nm), and blue (420 nm) illumination beams. The microcontroller, the LED driver, and the tri-color LED were all integrated on a single PCB, which was powered by a 5V-1A voltage adapter and communicated with the TFT PCB through the LED power signal.

The illumination light passes through the transparent solid agar and forms the lensfree images of the growing bacterial colonies on the TFT image sensor. The distance between the LED and the sample (i.e., the $z_1$ distance shown in Fig. 1(c)), is ~15.5 cm, which is large enough to make the illumination light uniformly cover the whole sample surface. The distance between the sample and the sensor ($z_2$) is roughly equal to the thickness of the solid agar, which is ~5 mm. The mechanical support material for the PCB, the sample, and the sensor were custom fabricated using a 3D printer (Objet30 Pro, Stratasys, Minnesota, USA).

**Image data acquisition**

Time-lapse imaging experiments were conducted to collect the data for both the training and testing phases. Our CFU imaging modality captured the time-lapse images of the agar plate under test every 5 min under red, green, and blue illuminations. A controlling program with a graphical user interface (GUI) was developed to perform the illumination switching and image capture automatically. The raw TFT images were saved in 12-bit format. After the experiments were completed, the samples were disposed of as solid biohazardous waste. In total, we collected the time-lapse TFT images of 889 *E. coli* colonies from 17 independent experiments to initially train our CFU detection neural network model. In addition to this, 442 bacterial colonies (128 *E. coli*, 126 *Citrobacter*, and 188 *K. pneumoniae*) were populated from 17 new agar plates and used to train (1) the final CFU detection neural network (through transfer learning from the initial detection model) and (2) the CFU classification neural network. A third independent dataset of 265 colonies from 13 new experiments was used to test the trained neural network models blindly.

**Bacterial colony candidate selection**

The entire candidate selection workflow consists of image pre-processing, differential analysis, colony mask segmentation, and candidate position localization, following the steps listed in Fig. 5(a-i). For each time point, three raw TFT images (red, green, and blue channels) were obtained over a FOV of ~10 cm². After getting the TFT images $I_{N\_raw, C}$, where $N$ refers to the $N$-th image obtained at $T_N$ and $C$ represents the color channels, R (red), G (green), and B (blue), a series of pre-processing operations were performed to enhance the image contrast. First, as shown in Fig. 5(a-b), the images were 5 times interpolated and normalized by directly subtracting the first frame at $T_0$. After this normalization step, the background regions had ~0 signal, while the regions representing the growing colonies had negative values because the colonies partially blocked and scattered the illumination light. Then, by adding 127 and saving the images as unsigned 8-bit integer arrays, the current frame at $T_N$ was scaled to 0-127, noted as $I_{N\_norm, C}$. Following the steps in Fig. 5 (b-c), $I_{N\_norm, C}$ was averaged as shown in Equation (1) to perform smoothing in the time domain, which yields $I_{N\_denoised, C}$:

$$I_{N\_denoised,C} = \frac{1}{3}\sum_{n=N-2}^{n=N} I_{n\_norm,C} \quad (1)$$

To further improve the sensitivity of our system, differential images $I_{N\_diff}$ averaged on three color channels were calculated as follows:

$$I_{N\_diff} = \frac{1}{3}\sum_{C=R,G,B}\left(I_{N\_denoised,C} - I_{(N-1)\_denoised,C}\right) \quad (2)$$

By this operation, the signals of static artifacts were suppressed, and the spatio-temporal signals of the growing colonies were enhanced as ring-shaped patterns. Next, a pixel-wise minimum intensity projection was performed, as shown in Fig. 5(e-f), to project the minimum intensity of the differential images from $I_{(N-7)\_diff}$ to $I_{N\_diff}$, yielding the image $I_{N\_projection}$. Following this step, with an empirically set intensity threshold, $I_{N\_projection}$ was segmented into a binary mask. After morphological operations to fill the ring-shaped patterns and a watershed-based[32] division of clustered regions, $M_N$ was obtained as presented in Fig. 5(g). Based on this binary mask, $M_N$, we extracted the connected components and localized their centroids as shown in Fig. 5(h). These centroid coordinates were dynamically updated for each time point to ensure maintaining the localization at the center of the growing colonies.

Despite this pre-processing of the acquired TFT images, there are still some time-varying non-colony objects that can be selected as false colony candidates (such as bubbles, dust, or other features created by the uncontrolled motion of the agar surface). Therefore, a deep neural network was trained to further screen each colony candidate to eliminate false positives, the details of which will be discussed in the next subsection.

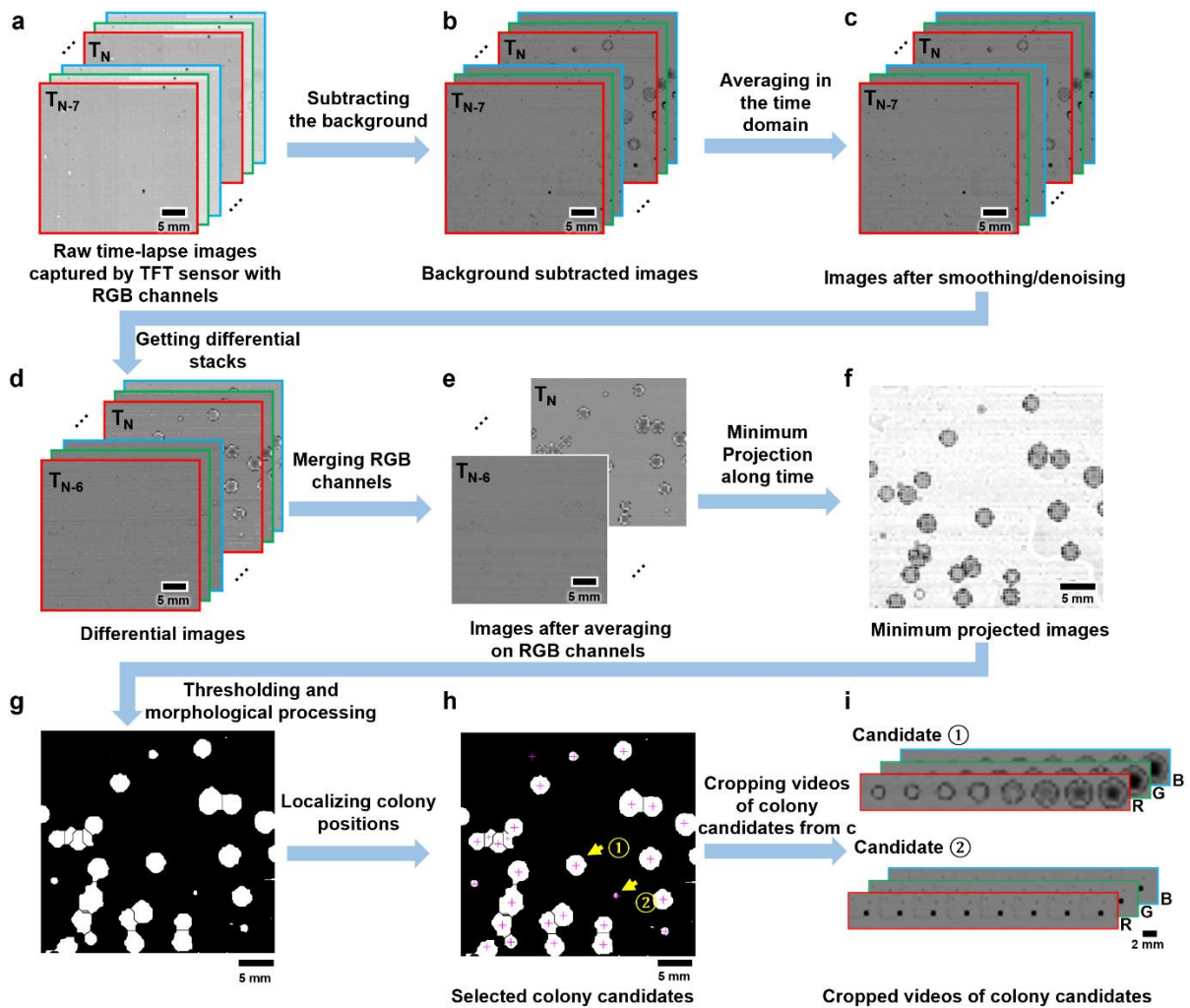

**Fig. 5: Bacterial colony candidate generation workflow.** The image pre-processing steps (a-i) were performed on the acquired TFT images in order to select the colony candidates; the cropped videos of the colony candidates were then passed through a trained CFU detection neural network to determine the true positives and eliminate false positives.

## DNN-based detection of bacterial colony growth

The time-lapse video of each colony candidate region across 8 frames of $I_{N\_denoised,\ C}$ was cropped as shown in Fig. 5. These videos were then up-sampled in the spatial domain and organized as a four-dimensional array ($3 \times 8 \times 160 \times 160$, i.e., color channels $\times$ number of frames $\times x \times y$) to be fed into the CFU detection neural network, which adopted the architecture of Dense-Net[33], but with 2D convolutional layers replaced by pseudo-3D convolutional layers[34] (see Fig. 6). The weights of this CFU detection DNN were initialized with a pre-trained model obtained on the *E. coli* CFU dataset with a single illumination wavelength of 515 nm. This pre-trained model was obtained using a total of 889 colonies (positives) and 159 non-colony objects (negatives) from 17 independent agar plates. Then, this initial neural network model was transferred to the multiple-species image dataset with multi-wavelength illumination, using 442 new colonies and 135 non-colony objects from another 17 independent agar plates. Both the positive image dataset and the negative image dataset were augmented across the time domain with different starting and ending time points, resulting in more than 10,000 videos used for training. A 5-fold cross-validation strategy was adopted to select the best hyper-parameter combinations. Once the hyper-parameters were decided, all the collected data were used for training to finalize our CFU detection neural network model. Data augmentation, such as flipping and rotation, was also applied when loading the training dataset.

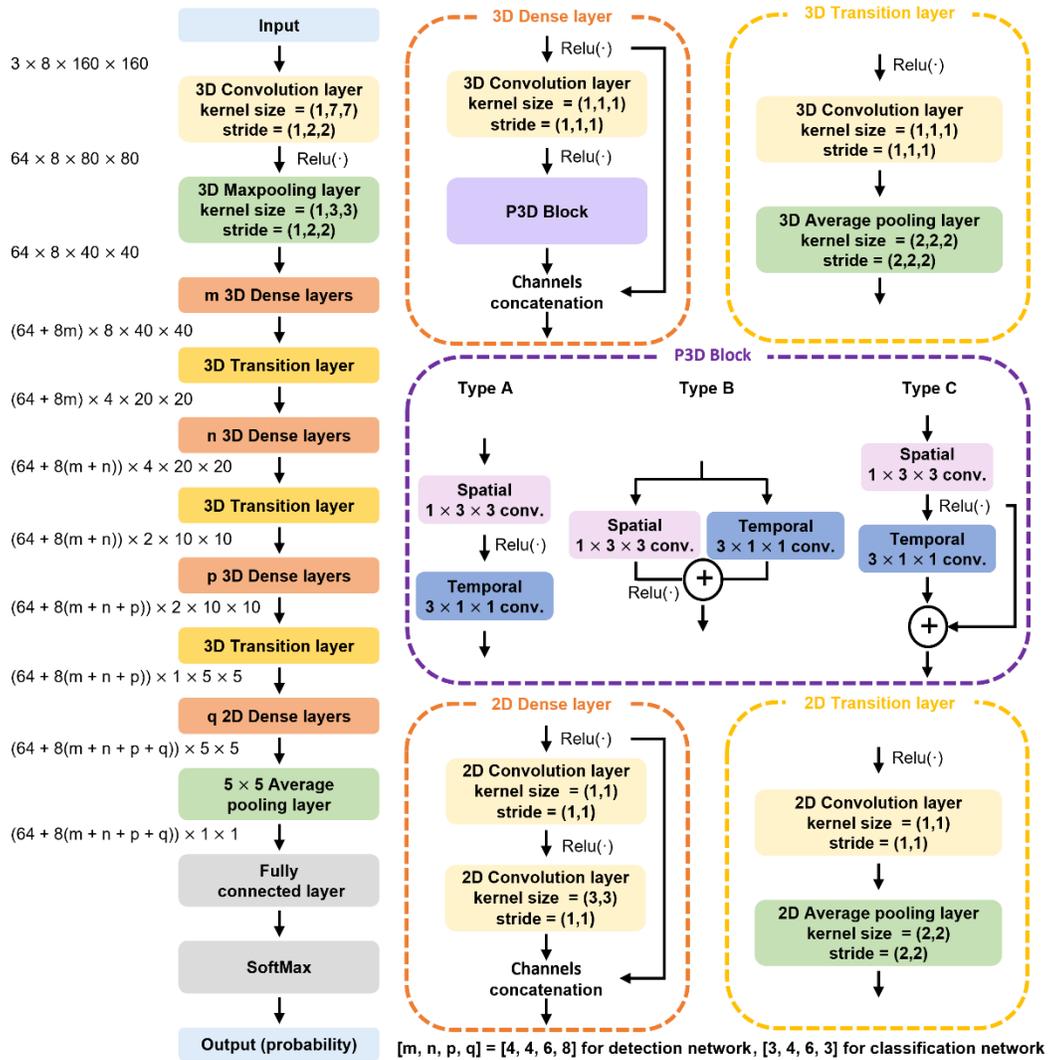

**Fig. 6: Network architectures for the CFU detection neural network and the CFU classification neural network.** A Dense-Net design was adopted here, with the 2D convolutional layers replaced by the pseudo-3D convolutional blocks. The CFU detection

and classification neural networks shared the same architecture, but the hyper-parameters [m, n, p, q] are selected to be different.

The network model was optimized using the Adam optimizer with a momentum coefficient of (0.9, 0.999). The learning rate started as $1\times10^{-4}$ and a scheduler was used to decrease the learning rate with a coefficient of 0.8 at every 10 epochs. The batch size was set to 8. The loss function was selected as:

$$l(p,g) = \sum_{k=1}^{k=K}\sum_{l=1}^{2} -w_l \cdot g_{k,l} \cdot \log\left(\frac{\exp(p_{k,l})}{\exp(p_{k,1}) + \exp(p_{k,2})}\right) \quad (3)$$

where $p$ is the network output, which is the probability of each class before the SoftMax layer, $g$ is the ground-truth label (which is equal to 0 or 1 for binary classification), $K$ is the total number of training samples in one batch, $w$ is the weight assigned to each class, defined as $w = 1 - d$ where $d$ is the percentage of the samples in one class. The training process was performed using a GPU (GTX1080Ti) which took ~5 hours to converge. With a decision threshold of 0.5, the CFU detection neural network converged with 92.6% sensitivity and 95.8% specificity. In the testing phase, the decision threshold was set to be 0.99, which achieved 100% specificity.

## DNN-based classification of *E. coli* and other total coliform colonies

To classify the species of the detected bacterial colonies, a second DNN-based classifier was built. The CFU classification neural network was trained on the same multi-wavelength dataset populated with 442 colonies (128 *E. coli* colonies, 126 *Citrobacter* colonies, and 188 *K. pneumonia* colonies). The input of the classification DNN was organized into a four-dimensional array (3×8×160×160, i.e., color channels×number of frames×$x$×$y$), but with a different normalization method. Different from the background subtraction normalization adopted for the CFU detection neural network, for the classification DNN, the network input was re-normalized by dividing the background intensities obtained at the first time point $T_0$. This division-based normalization was performed on three color channels so that the background would be normalized to ~1 in the three channels, revealing a white color in the background. Through this operation, the color variations across different experiments were minimized, improving the generalization capability of the classification DNN.

The network structure of the classification DNN was the same as the CFU detection network but with some differences in the hyper-parameter selection (see Fig. 6). The classification neural network model was initialized randomly and optimized using the Adam optimizer with a momentum coefficient of (0.9, 0.999). The learning rate started with $1\times10^{-3}$ and a scheduler was used to decrease the learning rate with a coefficient of 0.7 at every 30 epochs. The batch size was also set to 8. The classification neural network also used the weighted cross-entropy loss function as shown in Equation (3). The training process was performed using a GPU (GTX1080Ti) which took ~5 hours to converge. A decision threshold of 0.5 was used to classify the E. coli colonies and other total coliform colonies in the training process, achieving 91% and 97% accuracy, respectively. In the testing phase, the decision threshold was set to be 0.8, which achieved 100% classification accuracy. In addition, a colony size threshold of 4.5 mm$^2$ was used in the testing phase to ensure that only colonies that are large enough to identify their species were passed through the classification network.

## Supporting Information

Three supplementary videos (Videos S1-S3) are provided to exemplify the automated detection and classification performance for *E. coli*, *Citrobacter* and *K. pneumoniae* colonies.